%% file: hicss51.tex
\documentclass[10pt]{article}
\input{hicss51-packages.tex}
\newtheorem{Definition}{Definition}

\DeclareCaptionFont{9pt}{\fontsize{9pt}{10pt}\selectfont}
\title{Multi-View Community Detection in Facebook Public Pages}

\author{Zhige Xin \\
  Department of Computer Science \\
  University of California, Davis \\
  {\underline{zxin@ucdavis.edu}} \\
  \And
 Chun-Ming Lai \\
  Department of Computer Science \\
  University of California, Davis \\
  {\underline{cmlai@ucdavis.edu}} \\
   \And
  Jon W. Chapman \\
  Department of Computer Science \\
  University of California, Davis \\
  {\underline{jwchapman@ucdavis.edu}} \\
  \AND
  George Barnett \\
  Department of Communication \\
  University of California, Davis \\
  {\underline{gabarnett@ucdavis.edu}} \\
  \And 
  S. Felix Wu \\
  Department of Computer Science \\
  University of California, Davis \\
  {\underline{sfwu@ucdavis.edu}}\vspace{3.3cm}}

\date{}

\begin{document}
\maketitle
\begin{abstract}
Community detection in social networks is widely studied because of its
importance in uncovering how people connect and interact.
However, little attention has been given to community structure in Facebook public
pages.
In this study, we investigate the community detection problem in Facebook
newsgroup pages.
In particular, to deal with the diversity of user activities, we apply multi-view
clustering to integrate different views, for example, likes on posts and likes
on comments.
In this study, we explore the community structure in not only a given single page
but across multiple pages.
The results show that our method can effectively reduce isolates and improve the quality of community structure.

\end{abstract}

\section{Introduction}

In the last decade, the rapid growth and adoption of online social networks, such as Facebook, Twitter,
Linkedin, has fundamentally changed the way people interact with each other.
There are many people who would rather spend more time on these social networking sites than
traditional media.
With this trend, a great deal of data has been generated from the increasing
number of online social networking users.
Therefore, it is important to study the structure of social networks, which can
provide meaningful insight to Sociology, Communications, Economics, Marketing or even Epidemiology. 

One important type of structure of social networks is how the entities are divided into
different groups.
Basically, there is no formal definition of community, but it is believed   
that entities are densely connected inside each community with less links
between different communities \cite{wasserman1994social}.
This community structure plays a significant role in 
visualization \cite{kang2007visual}, dynamic community detection
\cite{tang2008community}, opinion mining \cite{wang2013influence}, and behavior
prediction \cite{tang2009scalable}. 

Previous research work on community detection generally dealt with the single-view setting.
Views are independent data sources or datasets.
One classic example is the web-page classification \cite{blum1998combining}, in
which one view is the content of web-page, and the other is comprised of the hyperlinks
pointing to it.
In social networks, such as YouTube and Flickr, the interactions between users are
complex \cite{tang2009uncovering}.
Similarly, on Facebook, users like, comment and share content, and
interact with each other through these activities.
Specifically, in the same page, the activities on posts and those on comments can
form two views. From each view, we can generate features and construct a graph to
find community structures within a given page.

Apart from the complexity of interactions of users, the volume of data
derived from social media has increased exponentially. 
At the end of 2017, Facebook had more than $2$ billion users
\cite{facebook2018} and up to $40$ million small businesses had public
pages \cite{page2018}.
We find that with an increase in the number of users, and corresponding
interactions, there is naturally a significant correlative increase in difficulty 
of discovering community structures within Facebook pages.
In this paper, we propose to model Facebook page as a weighted graph that is
generated by two views (posts and comments).
Then we examine the community structure of CBS News and The New Times
Facebook pages in last week of 2012.
In addition, the community structure for common users across multiple pages is
studied.
Our findings show that combining different views can remarkably reduce the
number of isolates in a single-view and make the community structure more
cohesive in networks because both views can mutually benefit from each other.

The rest of the paper is organized as follows.
The next section introduces the related work.
Section $3$ describes the issue of single-view methods.
The method is presented in Section $4$.
Section $5$ gives a detailed empirical study and results.
Finally, discussion is made in Section $6$.

\section{Related Work}

Before community detection became a trending research topic, data clustering has been always
a basic problem in machine learning research.
Although clustering is more general in terms of the non-overlapping
characteristic of data points, it can be applied to the community detection problem.
There are basically two types of clustering algorithms \cite{berkhin2006survey}.
One is model-based or so-called generative approach and the other is
similarity-based or discriminative approach. Furthermore, in similarity-model
methods, spectral clustering \cite{shi2000normalized,ng2002spectral} has prevailed in
the last several years for its performance, efficiency and robustness.
On the other hand, multi-view clustering is an effective tool for complex social networks.
The earliest work of multi-view clustering was proposed by Blum and Mitchell in
their co-training algorithm \cite{blum1998combining}.
Their idea was based on the assumption that learning can progress with enough
labeled data in each view otherwise each view mutually benefits from the labeled
data to each other.
Steffen Bickel and Tobias Scheffer \cite{bickel2004multi} proposed to alternately cluster
each view and exchange information during the learning process. 

Multi-view idea met spectral clustering in Virginia's work
\cite{de2005spectral}, in which the disagreement between two views was minimized.
It outperformed the result from each view in a sense that each view can leverage
the information from each other. Later, spectral clustering was extended to
clustering multiple graphs by Zhou et al \cite{zhou2007spectral}, where a
generalized Laplacian was built based on a random walk and its eigenvectors were computed.
Kumar et al. added co-training flavor to spectral clustering
\cite{kumar2011cotraining}.
First, original two sets of eigenvectors of graphs were computed.
Second, each eigenvector was used to modify the structure of the other graph.
Third, the clustering result was generated from the column-wise concatenation of the
two eigenvector matrices.
Also, Kumar et al. \cite{kumar2011coregularized} allowed the pair-wise co-regularizers
to be included in the objective function to decrease the disagreement between any pair of views.
Xiang et al. \cite{wang2013multi} were the first to bring the Pareto optimization into spectral
clustering to find optimal cuts for multiple graphs via multi-objective
functions.
No parameter was needed to set for Pareto frontier and it explored all possible good cuts.  
Xia et al. proposed a novel Markov chain method for Robust Multi-view Spectral
Clustering (RMSC) to deal with the noise problem \cite{xia2014robust}.
They used the Lagrangian Multiplier scheme to optimize RMSC. 
In addition to spectral clustering, many other clustering algorithms were applied to multi-view setting. Cai et al. extended K-means to multi-view clustering on a large scale amount of data \cite{cai2013multi}.
Linked Matrix Factorization \cite{tang2009clustering} was proposed for clustering multiple graphs.
And Liu et al. \cite{liu2013multi} formulated the problem as a joint non-negative matrix factorization process that forces each result of view to hold a consensus.
Zhang et al. \cite{zhang2015constrained} also used similar NMF method but with constrains to handle unmapped data.
Other related work includes Linked Matrix Factorization \cite{tang2009clustering}, tensor methods \cite{liu2013multiview}, fuzzy clustering \cite{jiang2015collaborative} and belief propagation \cite{wang2016multi}.  

Moreover, heterogeneous networks represent multi-relational in social networks
and some interesting work \cite{greene2009multi, tang2009uncovering,
  tang2012community} should be noticed recently.
However, little work has been carried out on Facebook public pages.
This paper focuses on modeling, testing and discovering the community
structure of Facebook newsgroup pages. To represent the users interactions
in public pages precisely, two graphs are built and merged in a way that
modularity is optimized.
Analysis is conducted on not only a single page but also against multi-page clusters.

\section{Issues of Single View Community Detection}

In the traditional research of community detection in the single-view setting, the
procedure is the following: first construct a graph based on the connection between
users and then apply some algorithm to partition the graph.
However, in real social networks, the interaction between users are complex.
For example, users can like or comment a post so it is hard to use only one
graph to represent different interactions between users. 

For CBS News page, we extract $205$ sample users from our database and
construct two graphs based on the interactions of different content: posts and comments.
The definition is shown in the next section.
Then, we apply the multi-level algorithm \cite{blondel2008fast,csardi2006igraph,hagberg2008exploring} to find
the community structure in the two graphs.
The connected nodes with the same color belong to the same community.

\begin{figure}[tbp]
\centering
\includegraphics[clip,width=0.9\linewidth]{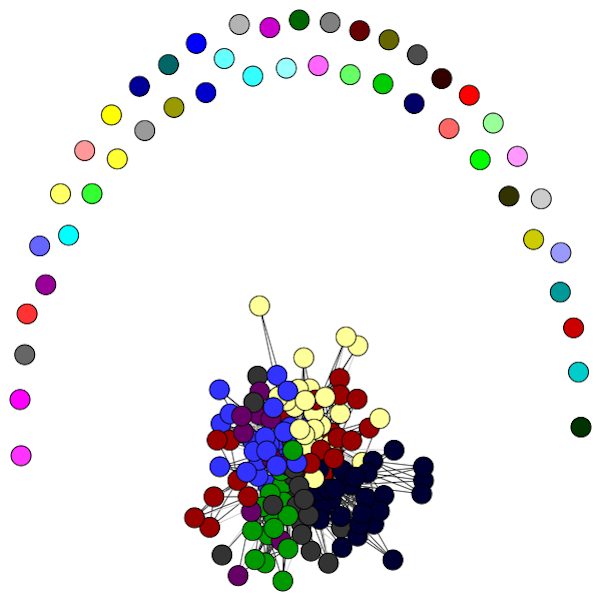}
\caption{The Community Structure of Post Graph for CBS}
\label{co_like_communities_cbs}
\end{figure}

\begin{figure}[tbp]
\centering
\includegraphics[clip,width=0.9\linewidth]{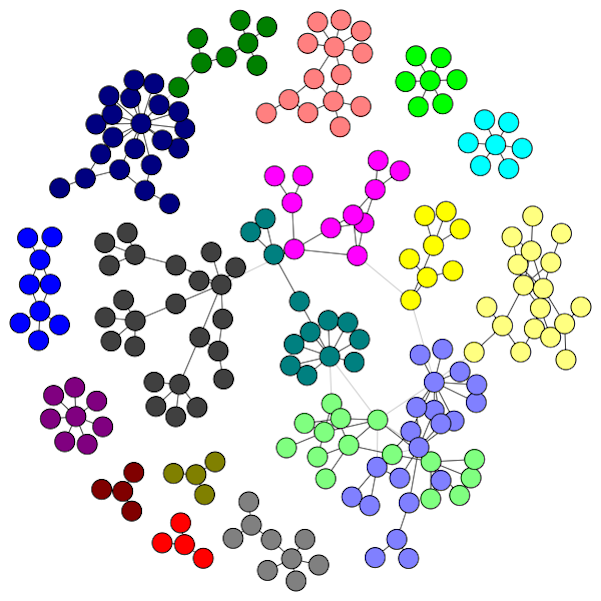}
\caption{The Community Structure of Comment Graph for CBS}
\label{like_comment_communities_cbs}
\end{figure}

It is clear that there are more isolates in the post graph in Figure \ref{co_like_communities_cbs} but the
comment graph in Figure \ref{like_comment_communities_cbs} shows a regular and balanced community structure.
Therefore, the issue is that single-view or one graph, can't best
represent the interactions between users.
In particular, in this example, the users that are considered as isolates in the
post graph actually have connections with others in the comment graph.

\section{Method}

\subsection{Problem Formulation}

Before we formulate the problem, it is necessary to mention the terminology in
this paper.
We consider networks as graphs, where a node represents a Facebook user and an edge
represents interaction between a pair of nodes.
Community and cluster are interchangeable as well. 

A multi-view dataset can be represented by \textbf{k} graphs that have the same
set of nodes but with a different set of edges.
Formally, given $m$ graphs $\mathcal{G}1=(\mathcal{V}1, \mathcal{E}1), \mathcal{G}2=(\mathcal{V}2,
\mathcal{E}2), ..., \mathcal{G}m=({\mathcal{V}m, \mathcal{E}m})$ and the number of communities
\textbf{k}, our goal is to find a vector \textbf{v}=($v_{1}$, $v_{2}$, ..., $v_{n}$) such that \textbf{v}
gives an optimal community structure for all graphs, where $v_{i}$ represents that node i belongs
to community $j$ and $1 \leq i \leq n$, $1 \leq j \leq k$.
In this paper, we focus on two views (activities on posts and comments) in
Facebook public pages.
Thus, $k = 2$.

\subsection{Graph Construction}

To some extent, community detection is a graph partitioning problem.
So, it is important to define the appropriate graph for our purpose.
In Facebook, users have three basic types of actions: comment, like and share.
Specifically, in newsgroup pages, post and comment are the basic blocks in which users interact with each other. 

In data clustering, a matrix is used to represent and analyze a graph.
Here we use adjacency matrices to represent our social interaction graphs.

If a pair of users $i$ and $j$ concurrently like a post, then we put $1$ in the cell $(i,
j)$ of the matrix.
And we call this the post graph.
The adjacency matrix \ref{adjacency1} for post view/graph is defined as follows:

\begin{equation}\label{adjacency1}
 A_{ij} = \left\{ \begin{array}{ll}
1 & \textrm{if \textit{i} and \textit{j} concurrently like the same post}\\
0 & \textrm{otherwise}\\ 
\end{array} \right.
\end{equation}

The other adjacency matrix \ref{adjacency2} for the comment view/graph is defined by likes on comment.
And we call this comment graph.
If user $i$ likes the comment of user $j$ or vice versa or they concurrently like a comment, we assign the
weight $B_{ij}$ to be 1, otherwise 0.

\begin{equation}\label{adjacency2}
 B_{ij} = \left\{ \begin{array}{ll}
                    1 & \textrm{if \textit{i} likes \textit{j}'s comment or vice versa or}\\
                      &  \textrm{\textit{i} and \textit{j} concurrently like a comment} \\
 0 & \textrm{otherwise} \\ 
\end{array} \right.
\end{equation}

Then we define our weighted graph by combining the two graphs into one and
assigning each graph a weight based on importance factor.
Moreover, it can be easily extended to multiple views.
Its formal definition \ref{adjacency} is as follows:

\begin{equation}\label{adjacency}
  W = \sum_{i=1}^{n}{\alpha_{i} X_{i}}
\end{equation}

where n is the number of views/graphs, $0 \leq \alpha \leq 1$ and $\sum_{i=1}^{n}{\alpha_{i}} = 1$.
And when $n=2$, it becomes the adjacency matrix for two views \ref{two_view}.

\begin{equation}\label{two_view}
  W = \alpha X_{1} + (1 - \alpha) X_{2}
\end{equation}

It turns out when $\alpha = 0$ or $\alpha = 1$, it is reduced to single-view.

\subsection{Multi-View Community Detection via Weighted Graphs}

To learn the optimal parameter in equation \ref{two_view}, modularity
\cite{newman2006modularity} is introduced.
Modularity \ref{modularity} is a measurement that evaluates
how apposite community structure is for any given network.
It ranges from $-1$ to $1$ inclusively and the larger it is, the better the
community structure is. 
From the definition, modularity essentially is the value that the real weight of
an edge minus the probability of generating it and sum them all.

\begin{equation}\label{modularity}
  Q=\frac{1}{2m}\sum_{ij}{\biggl[W_{ij}-\frac{d_{i}d_{j}}{2m}\biggr]}\delta(c_{i},c_{j})
\end{equation}

\begin{equation}
  \delta(c_{i},c_{j}) = \left\{ \begin{array}{ll}
                                  1 & \textrm{if \textit{i} and \textit{j} are in the same community}\\
0 & \textrm{otherwise}\\ 
\end{array} \right.
\end{equation}

where m is the number of edges, W is the adjacency matrix, $d_{i}, d_{j}$ are
the degree of node i and j respectively.

Our algorithm borrows the idea of modularity maximization \cite{chen2014community}.
First, we generate a set of parameters and calculate the modularity for each
network's structure.
Then, we pick the largest modularity value and its corresponding community structure
as the result.
The details can be seen in Algorithm \ref{weighted_algorithm}.

\begin{algorithm}
  \caption{Multi-View Community Detection via Weighted Graphs}
  \label{weighted_algorithm}
  \KwIn {Adjacency matrices $X_{i}$, and its parameter $\alpha_{i}$,
    where $\sum_{i=1}^{n}{\alpha_{i}} = 1$, $0 < \alpha_{i} < 1$ and $1 \leq i \leq n$}
  \KwOut {Indicator vector v}
  Initialize an empty vector set $V$ \;
  \ForEach {possible combination of $\alpha_{i}$}{
    generate the unified similarity matrix $W_{j}$ by \ref{adjacency};
    compute community structure $v_{j}$ using matrix $W_{j}$ \;
    put $v_{j}$ into $V$ \;
  }
  pick the $v$ in $V$ with largest modularity value as the final indicator vector \;
\end{algorithm}

The time complexity of Algorithm \ref{weighted_algorithm} is based on two aspects.
The first is to construct the adjacency matrix, which takes $O(n^{2})$ time
theoretically, but this can be reduced to $O(m)$ because most social networks
are sparse $(m \ll n^{2})$, where $n$ and $m$ are the number of nodes and edges of the network.
In addition, the second is to run the core community detection algorithm
\cite{blondel2008fast} $k$ times that takes $O(km)$, where $k$ is the number of parameters.
In total, the time complexity of our algorithm is $O(m+k*m)$, which is $O(m)$ since $k \ll m$.

\section{Empirical Study}

\subsection{Data Collection}

Our Facebook dataset \cite{erlandsson2015crawling} is crawled from public pages by using Facebook Graph API.
In this paper, we only consider the public pages like CBS News, Fox News and New
York Times, etc.
One characteristic of these pages is that only the administrators can post
information.
Users can only comment, like and share posts or reply and like comments.

In our database, we have several tables such as comment, likedby and shares, etc.
Since comment and like are the most common activities in Facebook newsgroup
pages, we focus on the comment table and the likedby table.
Specifically, the schema of comment is composed of $id$, $post$ $id$, $page$
$id$, $fb$ $id$, $message$, $can$ $remove$ and $created$ $time$.
And the schema of likedby is composed of $page$ $id$, $post$ $id$, $common$
$id$ and $fb$ $id$.

Until June 2018, we have crawled millions of pages, hundreds of millions of posts,
billions of comments and likes by near $10$ billions application-scoped users.
Table \ref{crawled} shows the statistics in detail.

\begin{table}[thb]
\centering
\small
\caption{The Metadata of Facebook Pages Crawled}
\label{crawled}
\begin{tabular}{lll}
\hline
  Table Name & Number of rows & Size in GB \\
\hline 
  page & 4,706,324 & 0.30 \\
  shares & 160,098,978 & 8.54 \\
  post & 343,000,084 & 246.66 \\
  comment & 6,429,663,271 & 1,213.59 \\
  likedby & 36,897,398,296 & 4,028.59 \\
  reaction & 95,831,789,680 & 9,973.33 \\
\hline
\end{tabular}
\end{table}

We collected two datasets from CBS News and The New York Times in the last week of
2012.
The statistics of the two datasets are listed in Table \ref{statistics_datasets}.

\begin{table}[thb]
\centering
\small
\caption{The Statistics of Facebook Newsgroups}
\label{statistics_datasets}
\begin{tabular}{lll}
\hline
  Category & CBS & NY Times \\
\hline 
  Users & 11,610 & 42,001 \\
  Posts & 42 & 57 \\
  Comments & 5,488 & 3,244 \\
  Likes & 15,000 & 64,104 \\
\hline
\end{tabular}
\end{table}

\begin{figure}[t]
\centering
    \subfloat[The User Distribution with Likes on Posts in CBS News Page]
    {\includegraphics[width=0.45\linewidth]{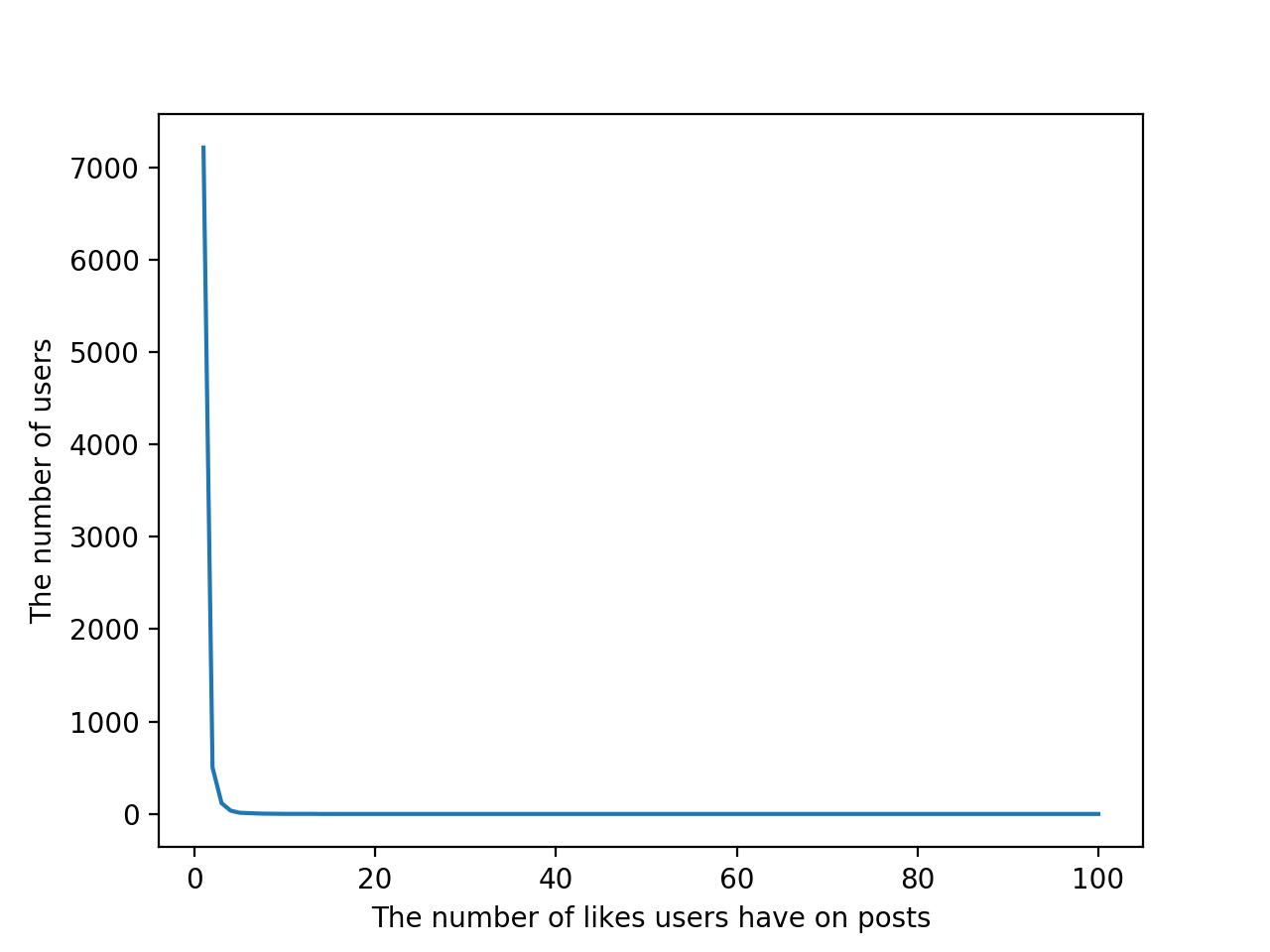}\label{1a}}
    \hfill
    \subfloat[The User Distribution with Likes on Posts in The New York Times Page]
    {\includegraphics[width=0.45\linewidth]{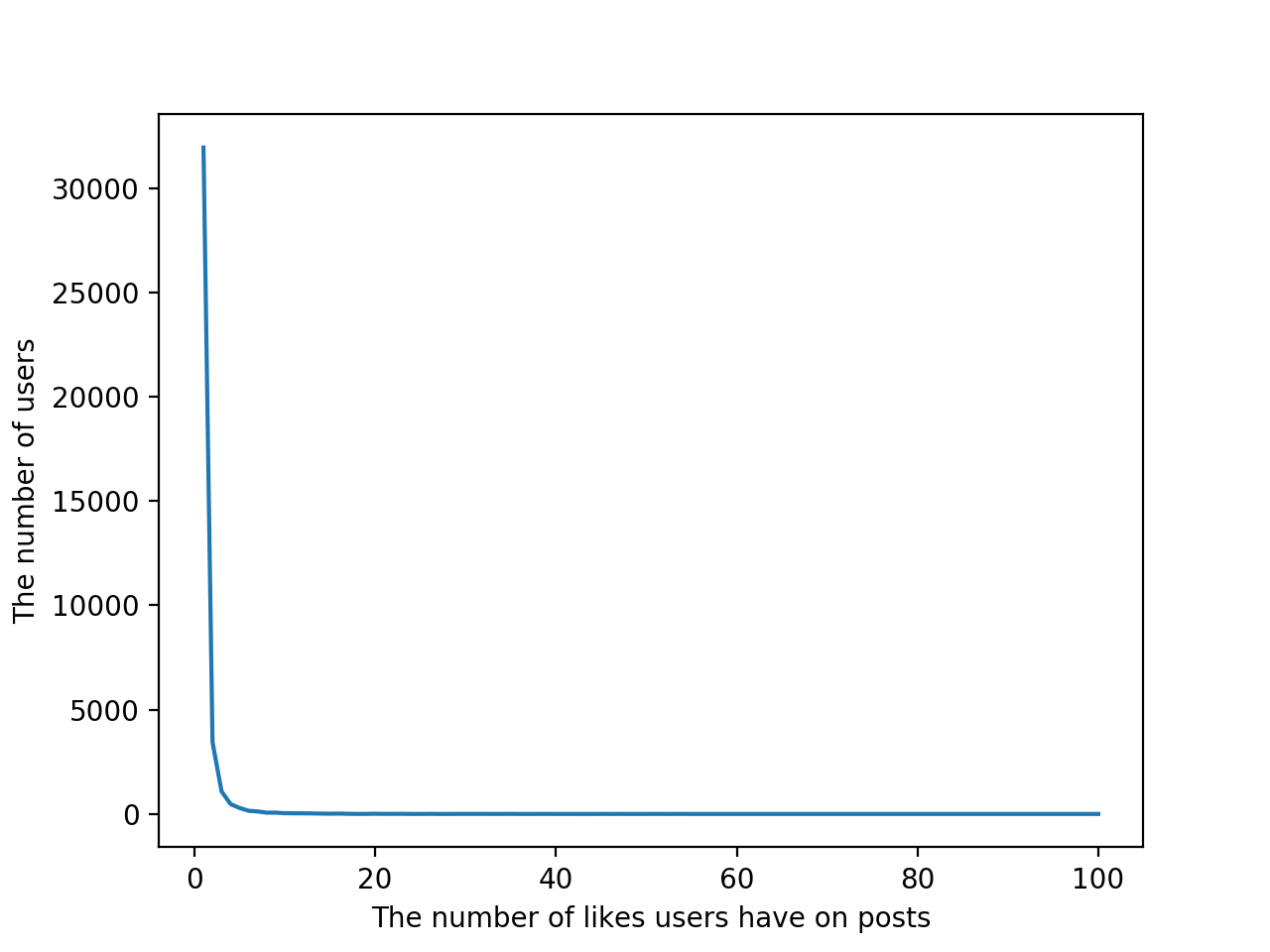}\label{1b}}
    \hfill
    \subfloat[The User Distribution with Likes on Comments in CBS News Page]
    {\includegraphics[width=0.45\linewidth]{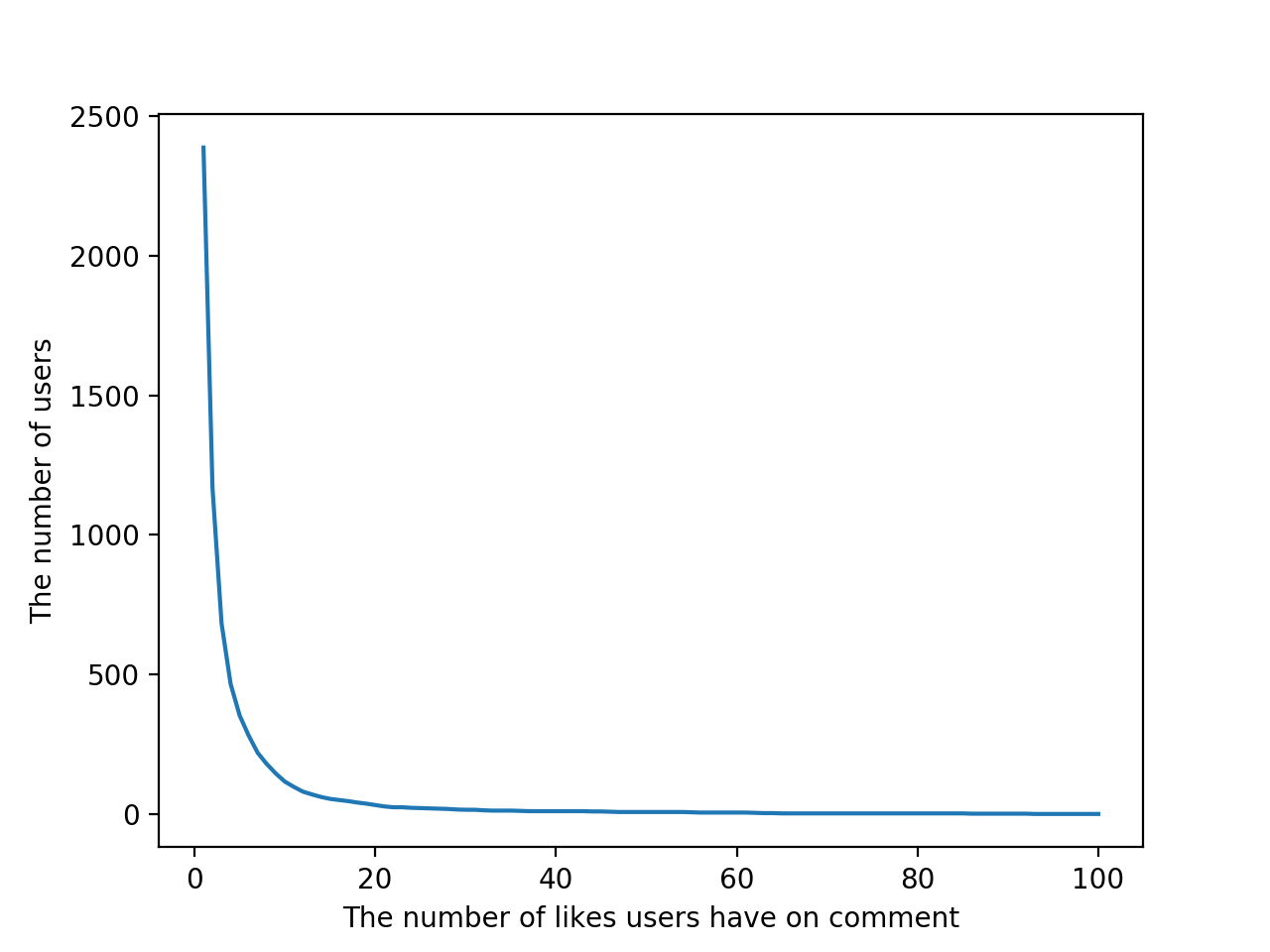}\label{2a}}
    \hfill
    \subfloat[The User Distribution with Likes on Comments in The New York Times
    Page]
    {\includegraphics[width=0.45\linewidth]{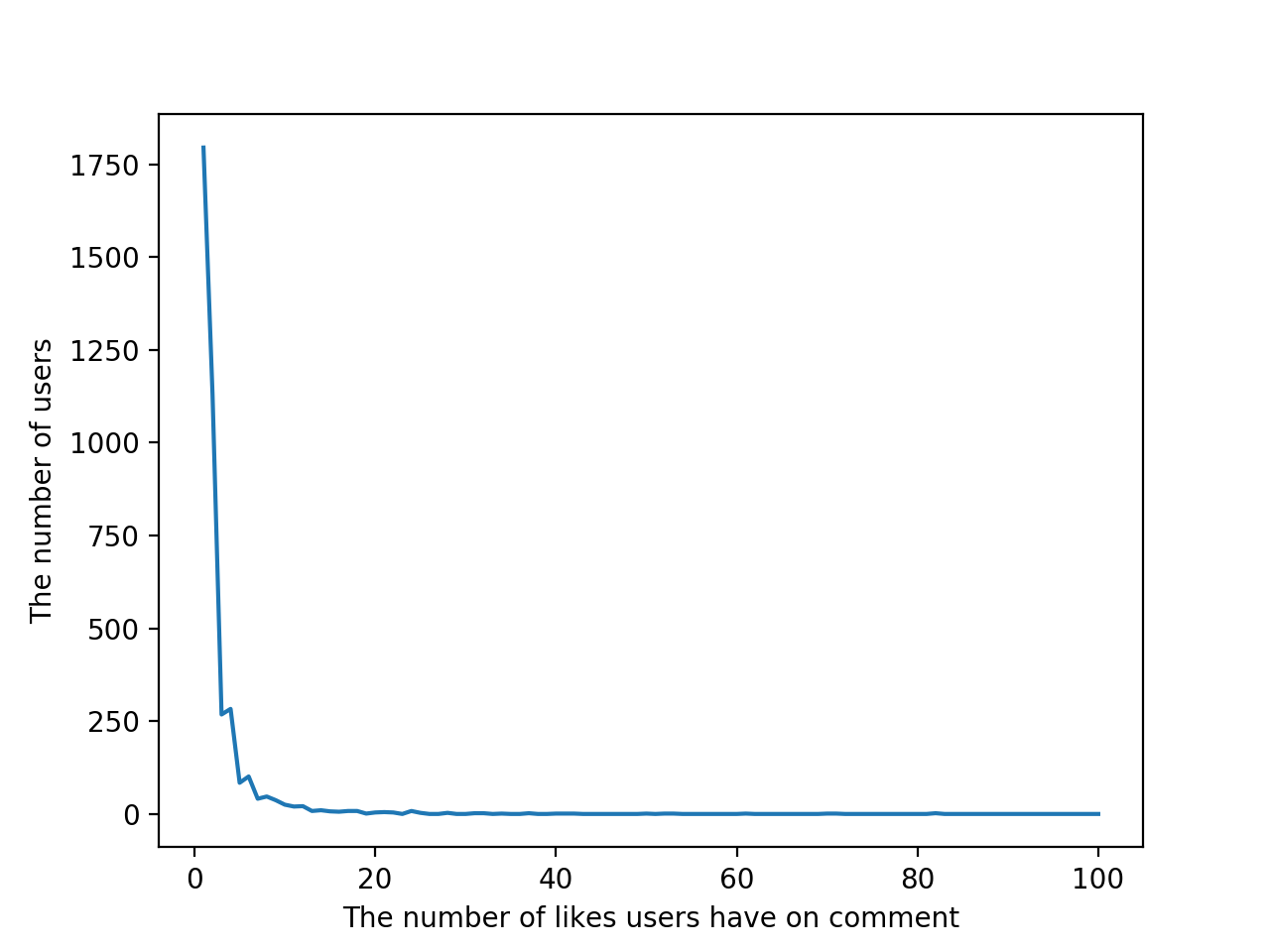}\label{2b}}
    \hfill
    \subfloat[The User Distribution with Comments in CBS News Page]
    {\includegraphics[width=0.45\linewidth]{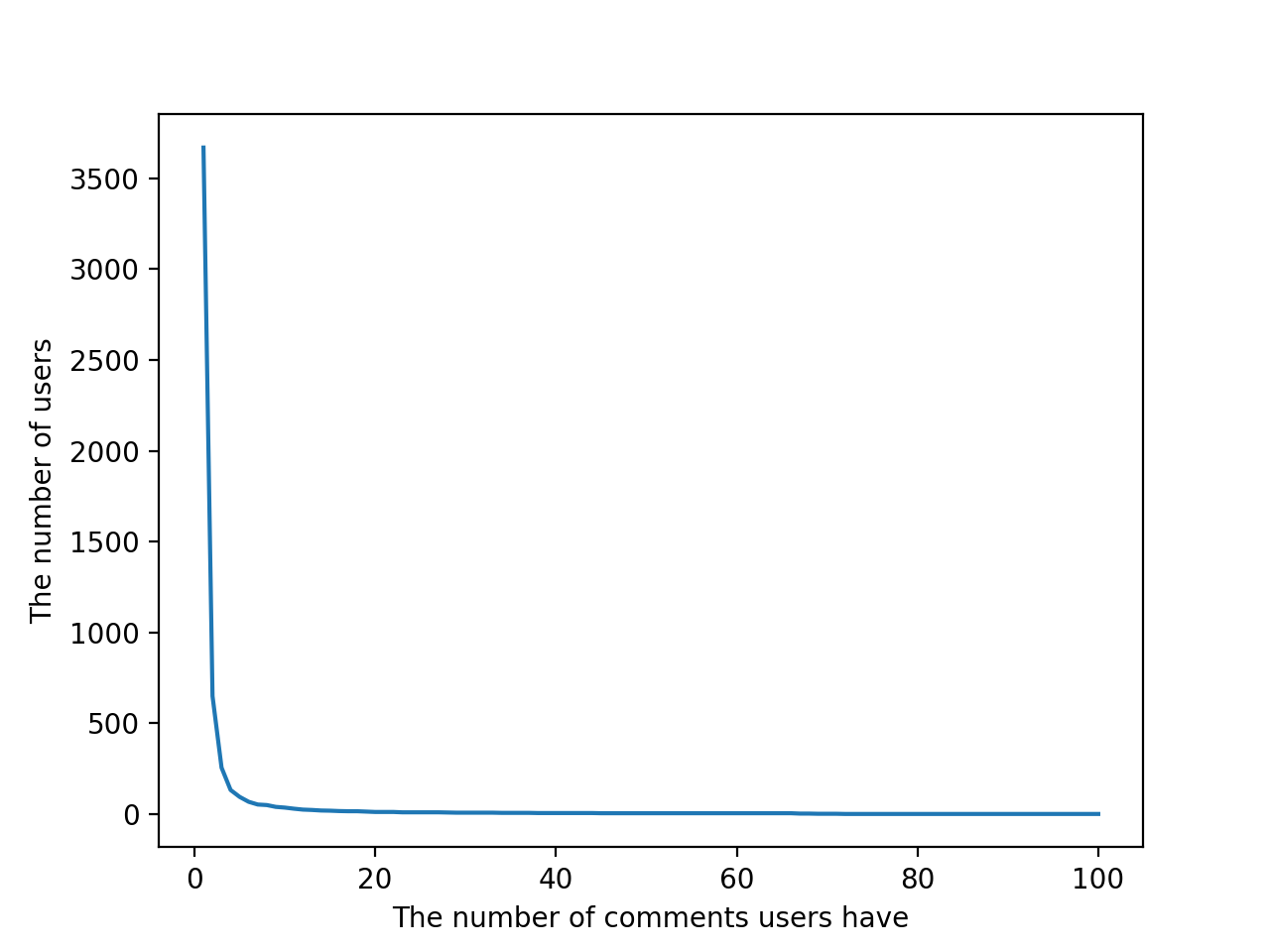}\label{3a}}
    \hfill
    \subfloat[The User Distribution with Comments in The New York Times Page]
    {\includegraphics[width=0.45\linewidth]{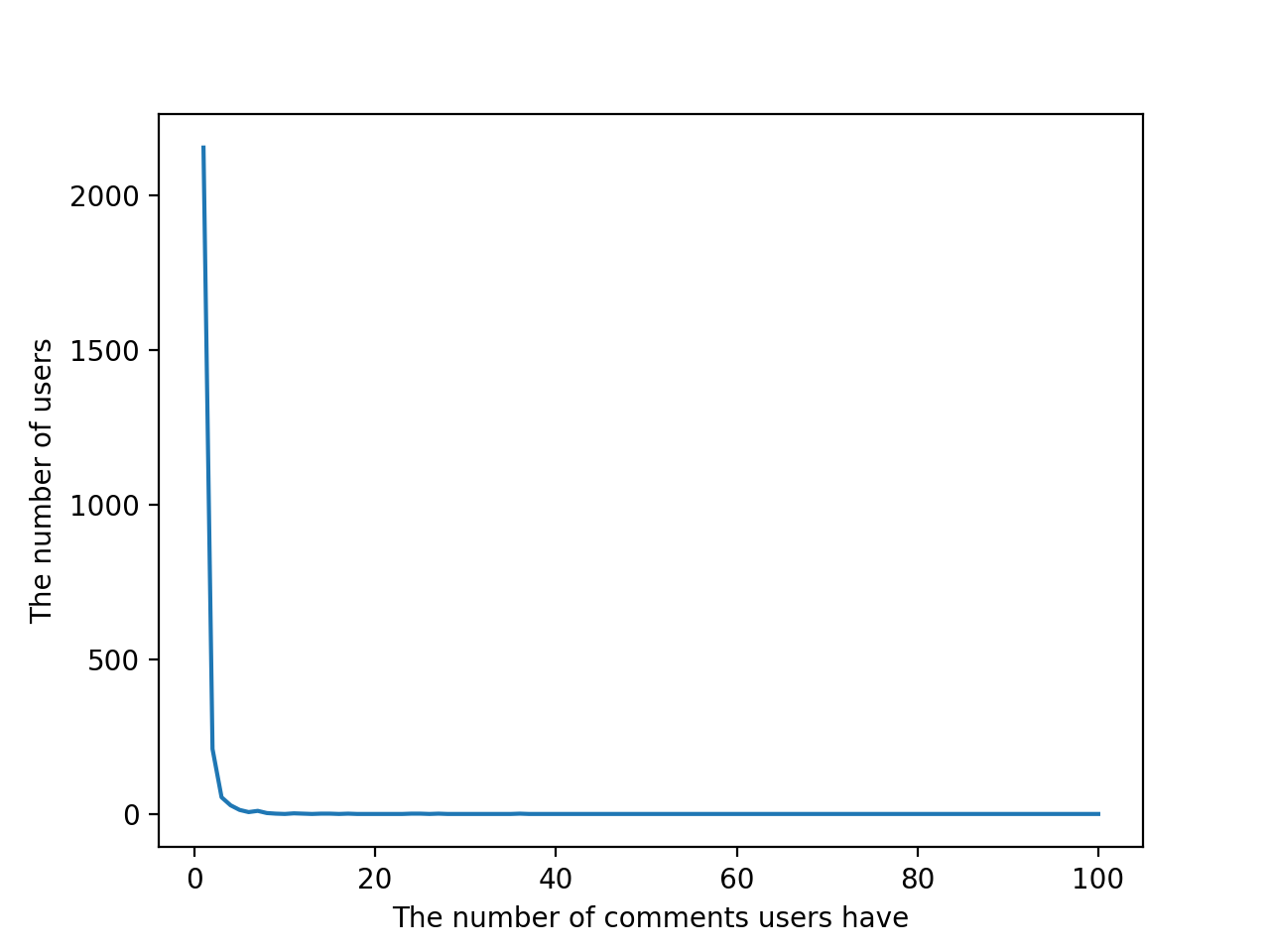}\label{3b}}
\caption{The User Distribution in CBS and NY Times}
\label{user_distribution}
\end{figure}

According to previous work \cite{faloutsos1999power, clauset2009power}, most variables in internet or social networks display power-law/long tail distributions.
To verify this, we plot the users distribution with the number of activities in
Facebook public pages, where the x-axis is the number of likes of users on comments or posts and the
y-axis is the number of users.
Figure \ref{user_distribution}\subref{1a} and Figure \ref{user_distribution}\subref{2a} reveal the users distributions with likes on
posts and on comments respectively for CBS News.
And Figure \ref{user_distribution}\subref{3a} plots the distributions of users with comments.
In addition, Figure \ref{user_distribution}\subref{1b}, Figure \ref{user_distribution}\subref{2b} and Figure \ref{user_distribution}\subref{3b} are those for
The New York Times.
It can be seen that with the increase in number of likes, the
number of users drops significantly, approximately power distributions.
Most Facebook users in the two pages have less than $10$ likes and $10$ comments in a week.

\subsection{Parameter Tuning}

After defining the graphs, we study how the parameter affects the community
structure of pages. 
Comment graphs show less density than post graphs, which implies that users are inclined to like posts rather than like comments.
Moreover, the number of clusters decrease tens of times in Table \ref{all_cbs}
and hundreds of times in Table \ref{all_nyt} for each page, which shows
our algorithm can effectively reduce isolates and uncover a more cohesive structure of networks.
View $1$ and View $2$ represent post graph and comment graph respectively.
Merged represents the weighted graph from the two views.
On the other hand, we observe that modularity is not a perfect measurement especially for networks
with a number of isolates because the more isolates are, the larger modularity is.

\begin{table}[thb]
\begin{center}
  \small
\caption{The Statistics of CBS News Graph} \label{all_cbs}
\begin{tabular}{cccc}
\hline
  Category & View 1 & View 2 & Merged \\
\hline
  Users & 10,535 & 10,535 & 10,535 \\
  Edges & 2,448,338 & 28,208 & 2,475,896 \\
  Clusters & 3,338 & 6,821 & \bf{120} \\
  Isolates & 3,321 & 6,647 & \bf{4} \\
  Modularity & 0.8334 & \bf{0.9135} & 0.8350 \\
\hline
\end{tabular}
\end{center}
\end{table}

\begin{table}[hb]
\begin{center}
  \small
\caption{The Statistics of The New York Times Graph} \label{all_nyt}
\begin{tabular}{cccc}
\hline
  Category & View 1 & View 2 & Merged \\
\hline
  Users & 41,252 & 41,252 & 41,252 \\
  Edges & 106,115,374 & 213,746 & 106,296,974 \\
  Clusters & 3,395 & 36,340 & \bf{31} \\
  Isolates & 3,383 & 36,279 & \bf{0} \\
  Modularity & 0.6054 & \bf{0.8057} & 0.6050 \\
\hline
\end{tabular}
\end{center}
\end{table}

\begin{figure}[t]
\centering
\includegraphics[clip,width=0.9\linewidth]{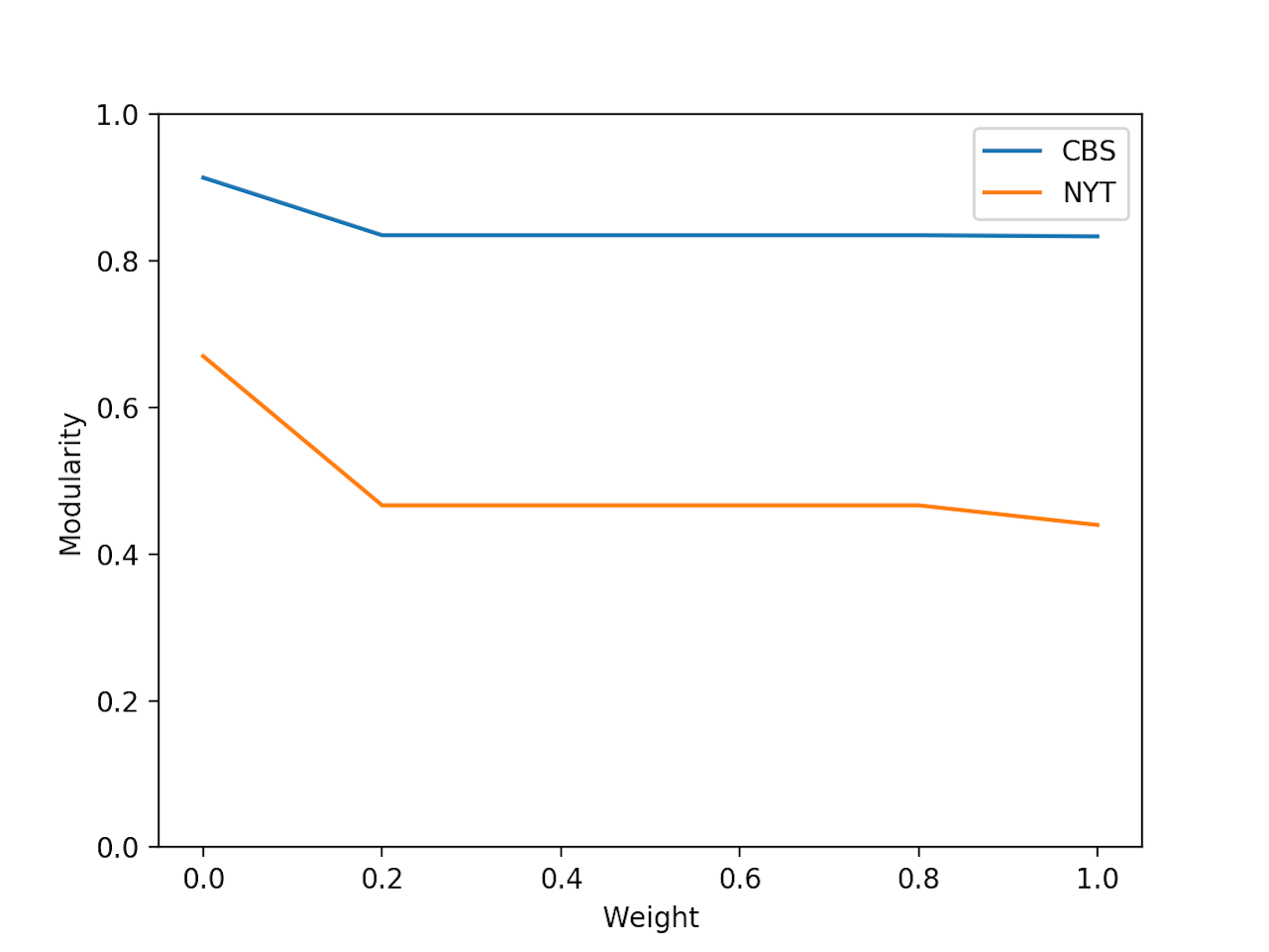}
\caption{The Relation between Weight and Modularity for CBS News and The New York Times}
\label{parameter_q}
\end{figure}

In order to find the optimal parameter, we generate a set of parameters as the candidates, \{0.0, 0.2, 0.4, 0.6, 0.8, 1.0\}.
Then we run Algorithm \ref{weighted_algorithm} repetitively for each parameter and calculate the corresponding modularity.
We plot the relation between the parameters and their modularity in Figure
\ref{parameter_q} for both CBS News and The New York Times.

Also we examine how the parameter affects the number of communities.
Table \ref{number_communities} shows the relation between the parameters and
their number of communities for the two pages.
The results demonstrate the parameter does not change the modularity value and
the number of communities.
In this study, the parameter value is to set to be $0.5$.
It is worth noting that the parameter improves the modularity of the post graph and it makes the both graphs denser.
Modularity is inclined to increase with more isolates for the same set of
users, which can explain why it reaches the maximum value when the parameter is $0$.

\begin{table}[thb]
\begin{center}
  \small
\caption{The Relation between Parameter and Number of Communities in CBS
  News and The New York Times} \label{number_communities}
\begin{tabular}{lll}
\hline
  Parameter & CBS & NY Times \\
\hline
  0 & 6,821 & 36,340\\
  0.2 & 120 & 31 \\
  0.4 & 120 & 31 \\
  0.6 & 120 & 31 \\
  0.8 & 120 & 31 \\
  1 & 3,338 & 3,395 \\
\hline
\end{tabular}
\end{center}
\end{table}

\subsection{Selection of Active Users}

Facebook itself is a huge dataset with billions of users so that it is
not easy to analyze the activities of all users. 
Moreover, there are a number of users who have limited activities and don't
contribute to providing any structural information.
Therefore, for our good, we define the following popular content and active users.
For popular content (post and comment) \ref{popular_content} in public pages is
defined by the following:

\begin{Definition}\label{popular_content}
Popular Facebook Content is a post liked by at least 2 users or a comment liked by at least 1 user.
\end{Definition}

For active Facebook users, the definition \ref{active_user} is as follows:

\begin{Definition}\label{active_user}
An Active Facebook User must have the following conditions: at least likes 1
popular post; at least has 1 popular comment or at least likes 1 popular comment.
\end{Definition}

For CBS News and The New York Times, we selected popular content and then the active users in the same page.
The result \ref{statistics_active} is shown in the following:

\begin{table}[thb]
\begin{center}
  \small
\caption{The Statistics of Active Users in Facebook Pages} \label{statistics_active}
\begin{tabular}{lll}
\hline
  Category & CBS & NY Times \\
\hline
  Users & 575 & 1,582 \\
  Posts & 42 & 54 \\
  Comments & 3,015 & 1,799 \\
  Likes & 15,000 & 64,014 \\
\hline
\end{tabular}
\end{center}
\end{table}

\subsection{Comparison of Different Methods}

We choose the multi-view clustering\cite{bickel2004multi} for the baseline.
To be fair, the weighted graph is taken as the base on which modularity is
calculated for all methods.
Multi-Level \cite{blondel2008fast} and LPA \cite{raghavan2007near} are selected
as our core community detection algorithms.
We tested these methods on the active users data described in Table \ref{statistics_active}.

Table \ref{cbs_comparison} and Table \ref{nyt_comparison} show the modularity
values calculated by a different method under the same graph and number of
communities.
We can see that our method based on the Multi-Level algorithm outperforms the other two in CBS
News and The New York Times pages.
On the other hand, like other clustering algorithms, MVC has to estimate the
optimal number of clusters so it repetitively runs at least K times, where K is
the maximum number of communities.

\begin{table}[thb]
\begin{center}
\caption{The Comparison of Methods for CBS} \label{cbs_comparison}
\begin{tabular}{ccc}
\hline
  Method & Modularity & Clusters \\
\hline
  MVC & 0.8196 & 14 \\
\hline
  Weighted LPA & 0.8200 & 26  \\
\hline
  Weighted Multi-Level & \bf{0.8373} & 20 \\
\hline
\end{tabular}
\end{center}
\end{table}

\begin{table}[thb]
\begin{center}
\caption{The Comparison of Methods for The New York Times} \label{nyt_comparison}
\begin{tabular}{ccc}
\hline
  Method & Modularity & Clusters \\
\hline
  MVC & 0.3899 & 7 \\
\hline
  Weighted LPA & 0.1641 & 4 \\
\hline
  Weighted Multi-Level & \bf{0.4241} & 10 \\
\hline
\end{tabular}
\end{center}
\end{table}

\subsection{Community Detection in Multiple Pages}

In reality, Facebook users often participate in multiple pages.
For example, one user can like a post about the presidential election in the CBS News page
and write a comment on the post with a similar topic on the NBC page.
Therefore, we try to discover communities of users across multiple pages in this part.

The first dataset is ABC News and CBS News in the December of 2012.
The reason why we chose the two pages is that they share enough Facebook users.
The statistics of the two pages can be seen in Table \ref{abc_cbs_stats}:

\begin{table}[hb]
\begin{center}
  \small
\caption{The Statistics of ABC and CBS in December 2012} \label{abc_cbs_stats}
\begin{tabular}{lll} 
\hline
  Category & ABC & CBS \\
\hline 
  Total users & 362,722 & 61,576 \\
  Common users & 961 & 961 \\
  Posts & 613 & 220 \\
  Comments & 108,116 & 24,564 \\
  Likes & 674,791 & 85,639 \\
\hline
\end{tabular}
\end{center}
\end{table}

\begin{figure}[tp]
\centering
\includegraphics[clip,width=0.9\linewidth]{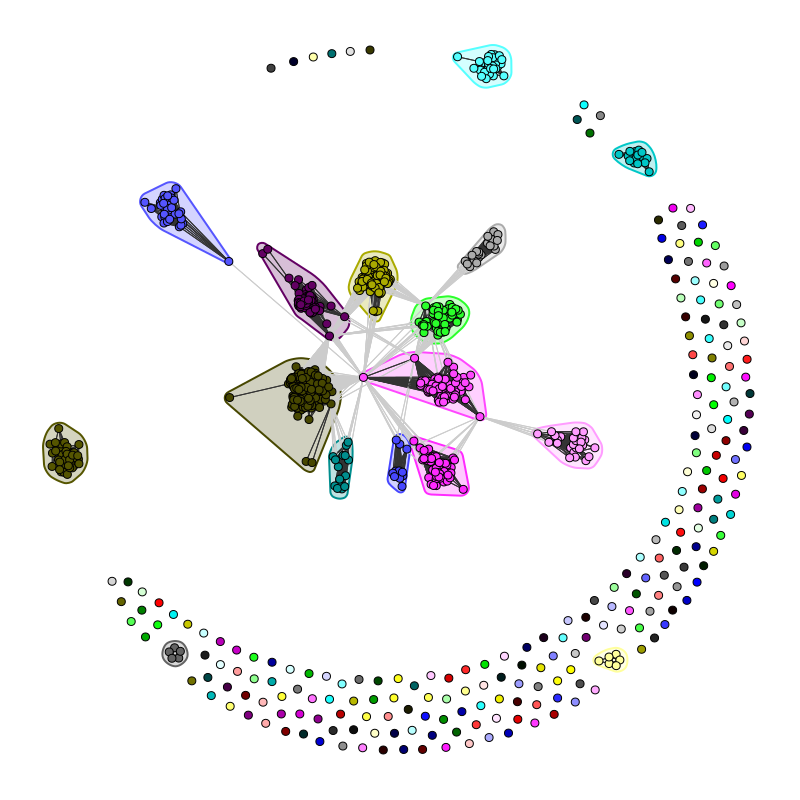}
\caption{The Community Structure of Post Graph for ABC and CBS}
\label{co_like_post_two}
\end{figure}

\begin{figure}[tp]
\centering
\includegraphics[clip,width=0.9\linewidth]{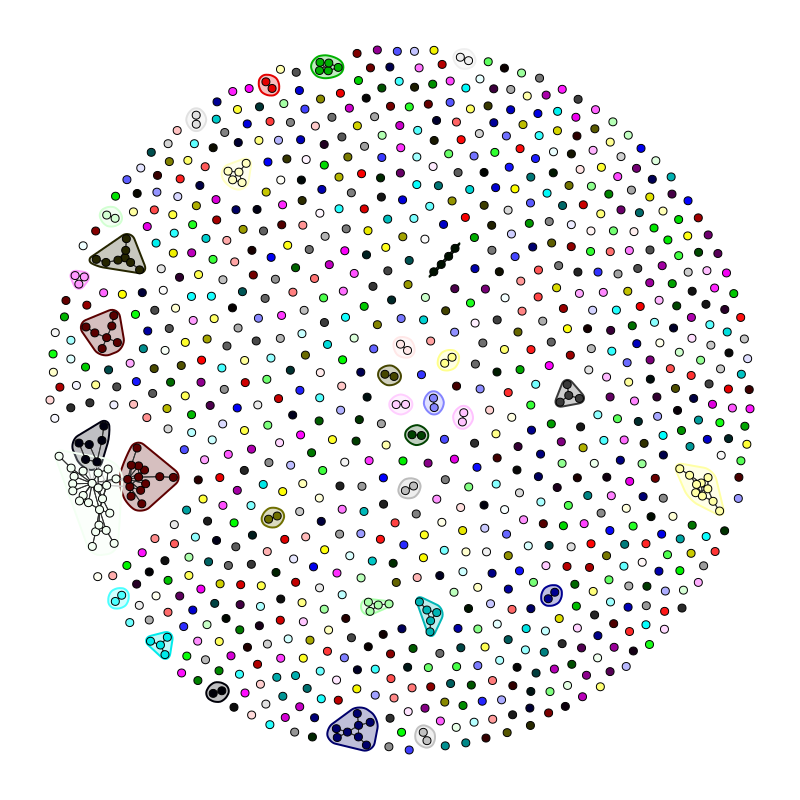}
\caption{The Community Structure of Comment Graph for ABC and CBS}
\label{comment_two}
\end{figure}

Here, $961$ common users who are involved in the two pages and let the parameter
$\alpha$ be $0.5$.
Initially, we plotted the post graph in Figure \ref{co_like_post_two} and the comment graph in Figure \ref{comment_two}.
By applying Algorithm \ref{weighted_algorithm}, we get the communities structure
for the graphs in Table \ref{abc_cbs_community}.

\begin{table}[thb]
\begin{center}
  \small
\caption{The Community Structure Information of ABC and CBS in December 2012} \label{abc_cbs_community}
\begin{tabular}{llll}
\hline
  Category & View 1 & View 2 & Merged \\
\hline
  Nodes & 961 & 961 & 961 \\
  Edges & 17,188 & 169 & 17,343 \\
  Clusters & 233 & 850 & 160 \\
  Isolates & 217 & 818 & \bf{137} \\
  Modularity & 0.6881 & \bf{0.8480} & 0.6912 \\
\hline
\end{tabular}
\end{center}
\end{table}

\begin{figure}[tp]
\centering
\includegraphics[clip,width=0.9\linewidth]{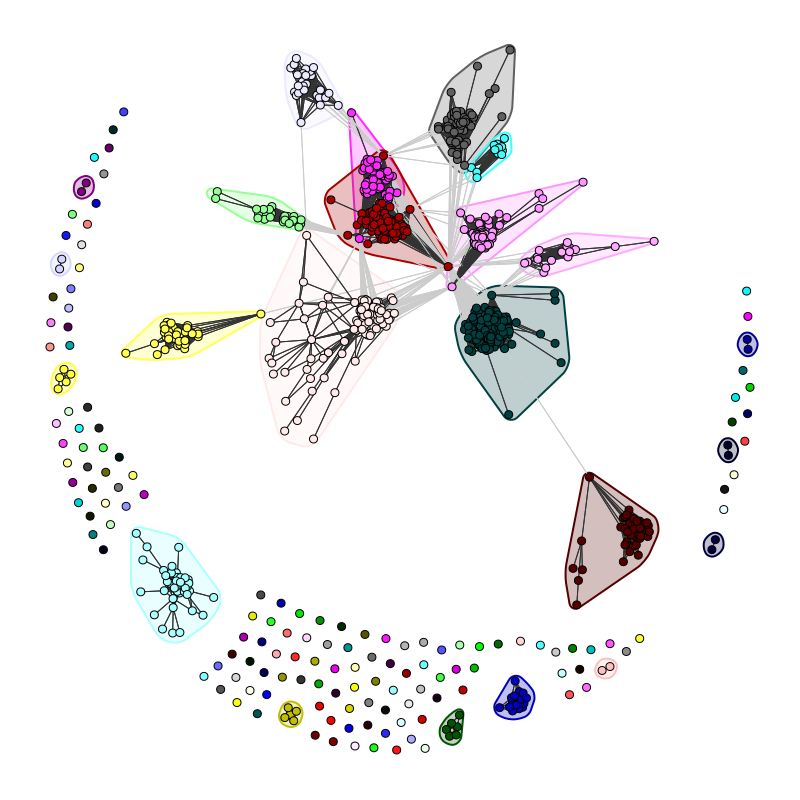}
\caption{The Community Structure of Weighted Graph in ABC and CBS}
\label{weighted_two}
\end{figure}

Both Figure \ref{co_like_post_two} and Figure \ref{comment_two} show sparse
structure. After we apply our multi-view community detection algorithm, it
\ref{weighted_two} erases most of the outliers in the comment graph.

The second dataset is ABC News, CBS News and NBC in the last half of $2012$.
In this dataset, we are interested in the users who have comment in ABC News and
likes in the rest two.
Consequently, we change the view $1$ into the graph in which any pair of users
concurrently like posts or comments and view $2$ into the graph in which any pair of users
like each other's comment.
The data statistics is shown in Table \ref{abc_cbs_nbc_data}. 
And the community structure information is described in Table \ref{abc_cbs_nbc_community}.

\begin{table}[thb]
\begin{center}
  \small
\caption{The Statistics of ABC, CBS and NBC of the Last Half of 2012} \label{abc_cbs_nbc_data}
\begin{tabular}{llll}
\hline
  Category & ABC & CBS & NBC \\
\hline
  Total users & 311,332 & 102,986 & 220,868 \\
  Common users & 533 & 533 & 533 \\
  Posts & 3,280 & 1,252 & 6,899 \\
  Comments & 588,708 & 160,097 & 486,728 \\
  Likes & 2,479,937 & 725,646 & 2,622,432 \\
\hline
\end{tabular}
\end{center}
\end{table}

\begin{table}[thb]
\begin{center}
  \small
\caption{The Community Structure Information of ABC, CBS and NBC in the Last Half
  2012} \label{abc_cbs_nbc_community}
\begin{tabular}{llll}
\hline
  Category & View 1 & View 2 & Merged \\
\hline
  Nodes & 533 & 533 & 533 \\
  Edges & 129,674 & 612 & 129,708 \\
  Clusters & 4 & 376 & \bf{6} \\
  Isolates & 1 & 361 & 1 \\
  Modularity & 0.0605 & \bf{0.5917} & 0.0604 \\
\hline
\end{tabular}
\end{center}
\end{table}

\begin{figure}[tbp]
\centering
\includegraphics[clip,width=0.9\linewidth]{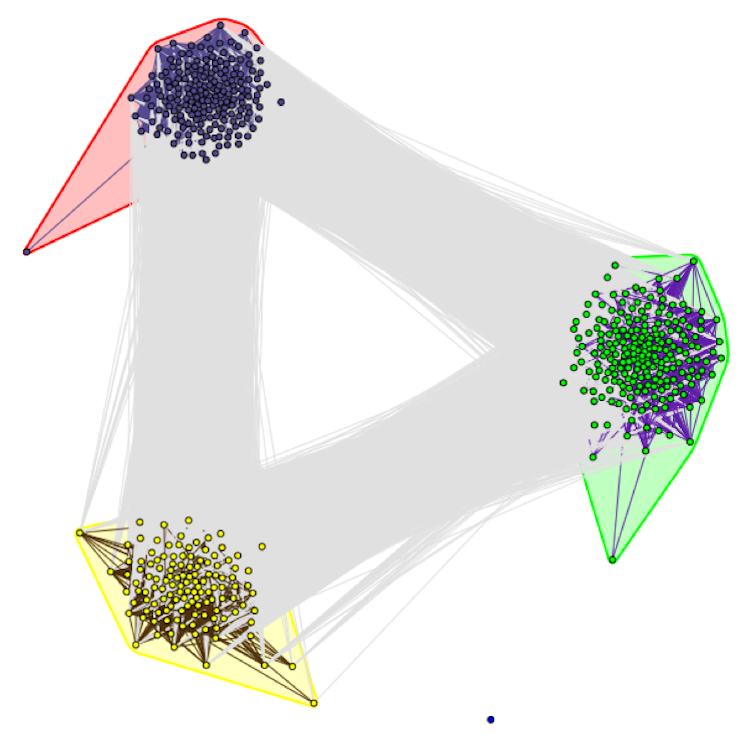}
\caption{The Community Structure of Post Graph for ABC, CBS and NBC}
\label{co_like_three}
\end{figure}

\begin{figure}[tbp]
\centering
\includegraphics[clip,width=0.9\linewidth]{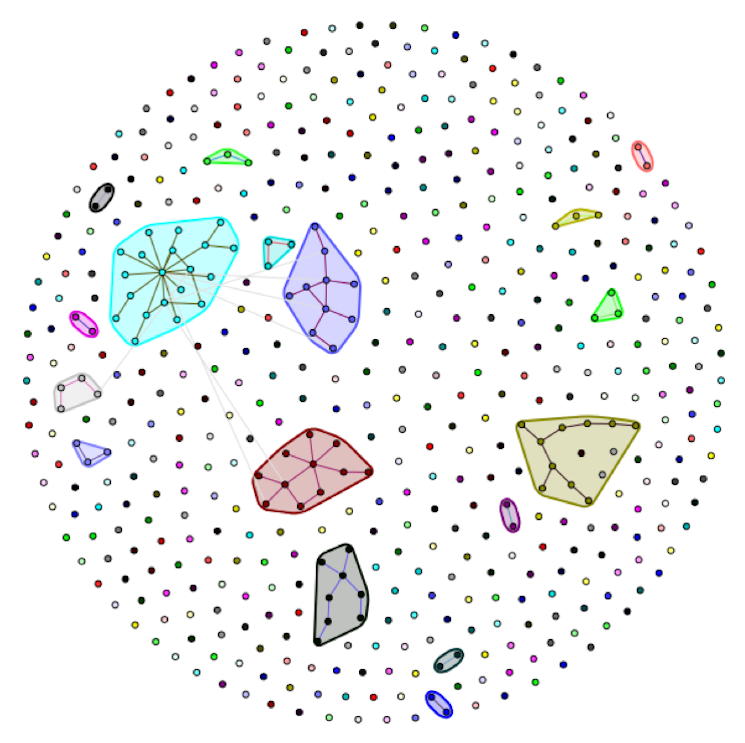}
\caption{ The Community Structure of Comment Graph for ABC, CBS and NBC}
\label{comment_three}
\end{figure}

\begin{figure}[tbp]
\centering
\includegraphics[clip,width=0.9\linewidth]{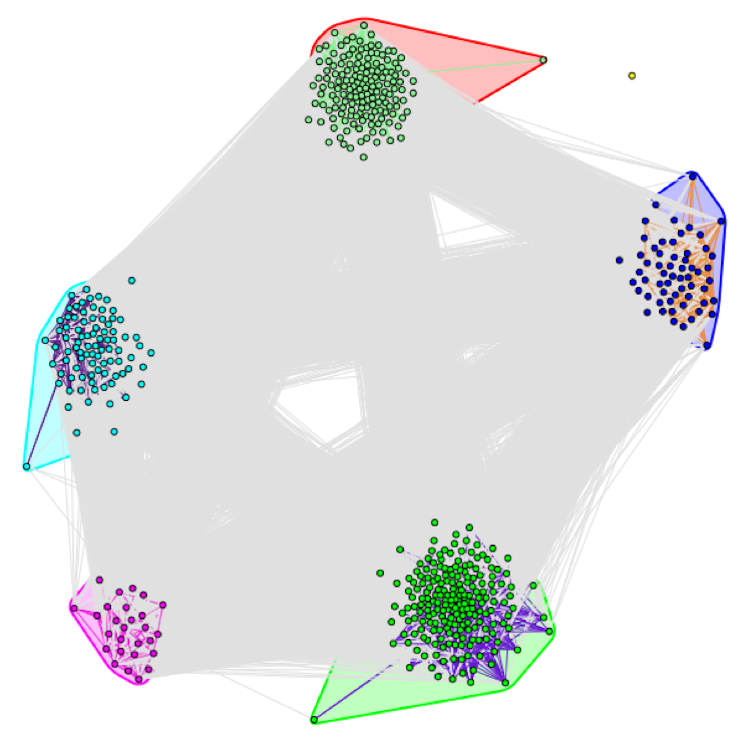}
\caption{The Community Structure of Weighted Graph for ABC, CBS and NBC}
\label{weighted_three}
\end{figure}

Figure \ref{co_like_three} illustrates the users are located in three communities but Figure \ref{comment_three} shows the sparse structure.
Figure \ref{weighted_three} describes that most isolates are effectively removed in the post graph.
Interestingly, when there are more than two pages, users within a page
interact more than different pages so each community represents each page.

\section{Discussion}

Community detection in social networks as a research topic has attracted a significant amount of attention in the past few years.
Prior research provided innovative theories, algorithms, and applications. However little work has been done with regards to exploring community structure in Facebook public
pages.
In this work, we propose a weighted multi-view community detection method and
apply it to the Facebook newsgroup pages, CBS News, and The New York Times, etc.
We not only investigate the community structure in a single page but across two
or even three pages.
The results reveal three advantages of our method:
1) it can alleviate the isolates issue in the sparse network/view.
For example, in the CBS News page (last week of 2012), the isolates are decreased from $3,321$ and $6,647$ to $4$.
2) more cohesive community structure can be found during the process, take the New
York Times page (last week of 2012) as an example, the number of communities
becomes $31$ from $3,395$ and $36,340$.
3) it discovers latent communities across multiple pages.
In the ABC, CBS and NBC pages, the common users show strong cohesion in post
graph, but in our weighted graph, two more communities are discovered, which might
provide useful information for recommend systems.

However, our method still has two limitations.
First, it can not handle directed graphs.
For example, like actions of users are one-way, which means one edge should
point from the user who has the like to the user who posts the comment. 
Under such situation, it is unknown whether the community structure could be
changed.
Second, overlapping communities can not be discovered, but it is interesting because it has 
more reasonable assumption that one user can belong to multiple communities.
More importantly, such overlapping structure might help to recommend users 
items more precisely in real recommendation systems.





\section{References} 


\bibliographystyle{ieeetr}
\bibliography{hicss51}

\end{document}

%% file: hicss51-packages.tex
\usepackage{hicss51}
\usepackage{times}
\usepackage[none]{hyphenat}
\usepackage{url}

\usepackage{latexsym}
\usepackage{indentfirst}
\usepackage{graphicx}
\usepackage{hyperref}
\usepackage{subfig}
\usepackage{array}
\usepackage{booktabs}
\usepackage{siunitx}
\usepackage[ruled]{algorithm2e}
\usepackage{amsmath}
\graphicspath{{images/}}